\documentclass[twocolumn,secnumarabic,amsmath,amssymb, nobibnotes, aps, prd]{revtex4-1}

\usepackage{graphicx}
\usepackage{subfigure}

\begin{document}

\title{Structure-preserving strategy for conservative simulation of relativistic nonlinear Landau--Fokker--Planck equation}%

\author{Takashi Shiroto}%
\email{shiroto-t@ile.osaka-u.ac.jp}
\affiliation{Institute of Laser Engineering, Osaka University, Suita, Osaka 565-0871, Japan}

\author{Yasuhiko Sentoku}%
\email{sentoku-y@ile.osaka-u.ac.jp}
\affiliation{Institute of Laser Engineering, Osaka University, Suita, Osaka 565-0871, Japan}

\begin{abstract}
Mathematical symmetries of the Beliaev--Budker kernel are the most important structure of the relativistic
Landau--Fokker--Planck equation. By preserving the beautiful symmetries, a mass-momentum-energy-conserving
simulation has been demonstrated without any artificial constraints.
\end{abstract}

\date{\today}%
\maketitle

\section{Introduction}

The Landau--Fokker--Planck (LFP) equation in the International System of Units is \cite{Landau--Lifshitz}

\begin{align}
\frac{\partial f_\mathrm{s}}{\partial t}=\frac{\Gamma}{2m_\mathrm{s}^2}
\frac{\partial}{\partial \mathbf{u}}\cdot \int \mathsf{U}(\mathbf{u},\mathbf{u'})
\cdot\left(f_\mathrm{s'}\frac{\partial f_\mathrm{s}}{\partial \mathbf{u}}
-\frac{m_\mathrm{s}}{m_\mathrm{s'}}f_\mathrm{s}\frac{\partial f_\mathrm{s'}}{\partial \mathbf{u'}}\right)
\mathrm{d}\mathbf{u'},\label{eq:1.1}
\end{align}
where $\Gamma=(q_\mathrm{s}^2 q_\mathrm{s'}^2 \log \Lambda)/4\pi {\varepsilon_0}^2$,
$\log \Lambda$ is the Coulomb logarithm, $\varepsilon_0$ is the vacuum permittivity,
$(f_\mathrm{s}, f_\mathrm{s'})$, $(m_\mathrm{s},m_\mathrm{s'})$, $(q_\mathrm{s},q_\mathrm{s'})$
and $(\mathbf{u}, \mathbf{u'})$ are the distribution function, mass, electric charge
and momentum per unit mass of species $(\mathrm{s},\mathrm{s'})$, respectively.
In the relativistic case, the collision kernel $\mathsf{U}(\mathbf{u},\mathbf{u'})$ is
described as follows \cite{Beliaev1956}:

\begin{align}
\mathsf{U}=\frac{r^2}{\gamma\gamma' w^3}
\left[w^2\mathsf{I}-\gamma^2\mathbf{vv}-\gamma'^2\mathbf{v'v'}+r\gamma\gamma'(\mathbf{vv'}+\mathbf{v'v})\right],\label{eq:1.2}
\end{align}
where $r=\gamma\gamma'(1-\mathbf{v}\cdot\mathbf{v'}/c^2)$, $w=c\sqrt{r^2-1}$,
$\mathbf{v}=\mathbf{u}/\gamma$, $\mathbf{v'}=\mathbf{u'}/\gamma'$, and $c$ is the speed of light in vacuum.
The Lorentz factor is defined as follows:

\begin{align}
\gamma=\sqrt{1+|\mathbf{u}/c|^2}=1/\sqrt{1-|\mathbf{v}/c|^2},\label{eq:1.3}\\
\gamma'=\sqrt{1+|\mathbf{u'}/c|^2}=1/\sqrt{1-|\mathbf{v'}/c|^2}.\label{eq:1.4}
\end{align}
The relativistic LFP equation is designed so as to ensure
the mass-momentum-energy conservation and the H-theorem.

The collisional processes described by the relativistic LFP equation are essential in fusion plasmas.
In practical simulations, the relativistic LFP equation is often linearized
so that the computational cost is reduced from $O(N^2)$ to $O(N)$,
where $N$ is the number of unknowns.
For fast ignition, which heats the imploded core by relativistic fast electrons, of inertial confinement fusion \cite{Tabak1994},
a linearization of Nakashima and Takabe \cite{Nakashima2002} is employed by relativistic Fokker--Planck codes
such as RFP-2D \cite{Yokota2006}, FIBMET \cite{Johzaki2009} and FIDO \cite{Sherlock2009,Ridgers2011}.
The linearization is based on the fact that the colliding particles are much
faster than the collided ones.
This violates symmetry of the collision kernel so the conservation laws are maintained only at the continuous limit.
For runaway electrons in tokamak disruptions \cite{Dreicer1959},
linearization which assumes an weakly relativistic equilibrium background \cite{Karney1985,Karney1986}
is sometimes performed to take into account the effect of non-thermal electrons \cite{Nuga2016},
while the conservation laws are violated.
Further, TASK/FP \cite{Nuga2011} and CQL3D \cite{Petrov2016} codes have options
that decompose the nonlinear LFP equation into Legendre modes and solve first a few modes.
The Legendre polynomials ensure the mass-momentum-energy conservation
limit to resolve structures of the pitch angle.
Recently, Stahl et al. developed NORSE code \cite{Stahl2017} which models nonlinear electron--electron collisions
by the Braams--Karney potential formulation \cite{Braams1987}.
However, the NORSE code violates the conservation laws including the mass conservation.
In the potential form, the ``nonlinear constraints'' is one of the ways to
preserve the conservation laws \cite{Taitano2015} and equilibrium state \cite{Taitano2017},
i.e. one of the projection method in the literature of applied mathematics \cite{Eich1993}.
However, the projection method can affect the stability of numerical schemes,
and it may not be suit for long time-scale simulations.
In magnetohydrodynamics simulations, for example, the projection method \cite{Brackbill1980}
or constrained transport (CT) method \cite{Evans1988}
has been used to enforce the solenoidal constraint ($\nabla\cdot\mathbf{B}=0$).
However, such an inconsistent magnetic field can induce ``checkerboard phenomenon''
which is one of the numerical instabilities \cite{Flock2010}.

Unlike the relativistic regime, many non-relativistic structure-preserving schemes have been proposed
to ensure the conservation laws, positivity and H-theorem.
Chang and Cooper developed a positivity-preserving scheme for one-dimensional (1D) linearized
LFP equation \cite{Chang1970}, and it was extended to
nonlinear isotropic LFP equation \cite{Buet2002} and
nonlinear multi-dimensional one \cite{Yoon2014,Yoon2014e,Hager2016} preserving the conservation laws and H-theorem.
A structure-preserving finite-element scheme \cite{Hirvijoki2017} is also developed
to ensure the conservation laws on unstructure meshes.
These works are based on a weak-form associated with Eq.~(\ref{eq:1.1}).
The integrand of Eq.~(\ref{eq:1.1}) can be transformed into the following one analytically:

\begin{align}
f_\mathrm{s}f_\mathrm{s'}\mathsf{U}(\mathbf{u},\mathbf{u'}) \cdot\left(\frac{\partial \log f_\mathrm{s}}{\partial \mathbf{u}}
-\frac{m_\mathrm{s}}{m_\mathrm{s'}}\frac{\partial \log f_\mathrm{s'}}{\partial \mathbf{u'}}\right).\label{eq:1.5}
\end{align}
The weak-form coming from Eq.~(\ref{eq:1.5}) is called as ``log'' weak-form,
and the proof of H-theorem by the log weak-form is straightforward.
An analytical discussion was first given by Pekker and Khudik \cite{Pekker1984},
and conservative and entropic discretizations have been proposed for
isotropic \cite{Berezin1987,Buet1998} and multi-dimensional cases \cite{Degond1994,Buet1997}.
Furthermore, an energy-conserving LFP scheme with the Rosenbluth potential form \cite{Rosenbluth1957}
was proposed using an analogy of the Maxwell stress tensor in electromagnetism \cite{Chacon2000-1,Chacon2000-2}.
However, structure-preserving schemes for the relativistic LFP equation have not been developed
since the Lorentz factor is not expressed as polynomials of finite-order in the momentum \cite{Hirvijoki2017}.

In this paper, we demonstrate a structure-preserving simulation of the relativistic LFP equation
which strictly preserve the conservation laws of mass, momentum and energy.
The key concept is the same with our recent work about a quadratic conservative scheme for the relativistic
Vlasov--Maxwell system \cite{Shiroto2019}.
Moreover, a similar approach has been used in a non-relativistic scheme with non-uniform meshes \cite{Buet2006,Buet2007}.
The rest of this paper is composed as follows.
Section~\ref{sec:2} shows an intuitive discretization which cannot maintain the energy conservation.
Section~\ref{sec:3} deductively derives requirements for the conservation laws in discrete form.
Our structure-preserving scheme and its concept are introduced in Sec.~\ref{sec:4}.
A verification of conservation property through a thermal-equilibration is performed in Sec.~\ref{sec:5}.
Section~\ref{sec:6} is conclusions of this article.

\section{\label{sec:2}CONVENTIONAL SCHEME}

In this article, the relativistic LFP equation is discretized as follows.
Note that the conservation laws are not affected by the temporal structure, so only momentum dimensions are discretized here:

\begin{align}
\frac{\partial {f_\mathrm{s}}^\mathbf{j}}{\partial t}\equiv\frac{\Gamma}{2{m_\mathrm{s}}^2}\frac{\delta}{\delta \mathbf{u}}\cdot
\sum_\mathbf{k} &\mathsf{U}(\mathbf{u^j},\mathbf{{u'}^k})\cdot\notag\\
&\left({f_\mathrm{s'}}^\mathbf{k} \frac{\delta {f_\mathrm{s}}^\mathbf{j}}{\delta \mathbf{u}}
-\frac{m_\mathrm{s}}{m_\mathrm{s'}}{f_\mathrm{s}}^\mathbf{j} \frac{\delta {f_\mathrm{s'}}^\mathbf{k}}{\delta \mathbf{u'}}
\right)\Delta \mathbf{u'},\label{eq:2.1}
\end{align}
where $\mathbf{j}\equiv[j_1,j_2,j_3]$ and $\mathbf{k}\equiv[k_1,k_2,k_3]$ are the indices of uniform momentum grids
for $\mathbf{u}$ and $\mathbf{u'}$, respectively.
$\Delta \mathbf{u}\equiv \Delta u_1 \Delta u_2 \Delta u_3$, and $(\Delta u_1, \Delta u_2, \Delta u_3)$ is
the grid interval of $(u_1, u_2, u_3)$.
The species $(\mathrm{s},\mathrm{s'})$ use the same momentum grids
$(\mathbf{u^j}=\mathbf{{u'}^j}, \Delta \mathbf{u}=\Delta \mathbf{u'})$.
Here we use a second-order central difference for simplicity:

\begin{align}
\frac{\delta a^{\mathbf{j},\mathbf{k}}}{\delta u_1}\equiv
\frac{a^{j_1+1,j_2,j_3,k_1,k_2,k_3}-a^{j_1-1,j_2,j_3,k_1,k_2,k_3}}{2\Delta u_1},\notag\\
\frac{\delta a^{\mathbf{j},\mathbf{k}}}{\delta u_2}\equiv
\frac{a^{j_1,j_2+1,j_3,k_1,k_2,k_3}-a^{j_1,j_2-1,j_3,k_1,k_2,k_3}}{2\Delta u_2},\notag\\
\frac{\delta a^{\mathbf{j},\mathbf{k}}}{\delta u_3}\equiv
\frac{a^{j_1,j_2,j_3+1,k_1,k_2,k_3}-a^{j_1,j_2,j_3-1,k_1,k_2,k_3}}{2\Delta u_3},\notag
\end{align}
\begin{align}
\frac{\delta a^{\mathbf{j},\mathbf{k}}}{\delta u'_1}\equiv
\frac{a^{j_1,j_2,j_3,k_1+1,k_2,k_3}-a^{j_1,j_2,j_3,k_1-1,k_2,k_3}}{2\Delta u'_1},\notag\\
\frac{\delta a^{\mathbf{j},\mathbf{k}}}{\delta u'_2}\equiv
\frac{a^{j_1,j_2,j_3,k_1,k_2+1,k_3}-a^{j_1,j_2,j_3,k_1,k_2-1,k_3}}{2\Delta u'_2},\notag\\
\frac{\delta a^{\mathbf{j},\mathbf{k}}}{\delta u'_3}\equiv
\frac{a^{j_1,j_2,j_3,k_1,k_2,k_3+1}-a^{j_1,j_2,j_3,k_1,k_2,k_3-1}}{2\Delta u'_3}.\notag
\end{align}
where $a^{\mathbf{j},\mathbf{k}}\equiv a(\mathbf{u^j},\mathbf{u'^k})$ is an arbitrary function.

In an intuitive discretization, the collision kernel is calculated by its arguments directly:

\begin{align}
\mathsf{U}(\mathbf{u^j},\mathbf{u'}^\mathbf{k})\equiv\frac{(r^\mathbf{j,k})^2}{\gamma^\mathbf{j}{\gamma'}^\mathbf{k} (w^\mathbf{j,k})^3}
\left[(w^\mathbf{j,k})^2\mathsf{I}-\mathbf{u^ju^j}-\mathbf{{u'}^k{u'}^k}\right.\notag\\
\left.+r^\mathbf{j,k}(\mathbf{u^j{u'}^k}+\mathbf{{u'}^ku^j})\right],
\label{eq:2.2}
\end{align}
From the definition, the collision kernel Eq.~(\ref{eq:2.2}) satisfies two mathematical symmetries:

\begin{align}
\mathsf{U}(\mathbf{u^j,{u'}^k})=\mathsf{U}(\mathbf{{u'}^k,u^j}),\label{eq:2.3}\\
\mathsf{U}(\mathbf{u^j},\mathbf{{u'}^k})\cdot\mathbf{v^j}=\mathsf{U}(\mathbf{{u'}^k},\mathbf{u^j})\cdot\mathbf{{v'}^k}.\label{eq:2.4}
\end{align}

\section{\label{sec:3}REQUIREMENTS FOR CONSERVATION LAWS IN DISCRETE FORM}
In this section, requirements for the conservation laws are derived deductively.
The mass conservation is trivially maintained,
so the discussion is focused on the momentum-energy conservation.
The following points are required to prove the conservation laws analytically from the relativistic LFP equation:
\begin{enumerate}
   \item Integration-by-parts must be maintained.
   \item $\mathsf{U}(\mathbf{u},\mathbf{u'})=\mathsf{U}(\mathbf{u'},\mathbf{u})$ is required for the momentum conservation.
   \item $\mathsf{U}(\mathbf{u},\mathbf{u'})\cdot\mathbf{v}=\mathsf{U}(\mathbf{u'},\mathbf{u})\cdot\mathbf{v'}$
         is required for the energy conservation.
\end{enumerate}
If one can assume $a^\mathbf{j,k}=0$ at the momentum boundaries,
the following summation-by-parts, i.e. the integration-by-parts in discrete form, is valid:

\begin{align}
\sum_\mathbf{j,k} \left(\frac{\delta a^\mathbf{j,k}}{\delta \mathbf{u}}b^\mathbf{j,k}\right)\Delta \mathbf{u}\Delta \mathbf{u'}
=-\sum_\mathbf{j,k} \left(a^\mathbf{j,k}\frac{\delta b^\mathbf{j,k}}{\delta \mathbf{u}}\right)\Delta \mathbf{u}\Delta \mathbf{u'},
\notag\\
\sum_\mathbf{j,k} \left(\frac{\delta a^\mathbf{j,k}}{\delta \mathbf{u'}}b^\mathbf{j,k}\right)\Delta \mathbf{u}\Delta \mathbf{u'}
=-\sum_\mathbf{j,k} \left(a^\mathbf{j,k}\frac{\delta b^\mathbf{j,k}}{\delta \mathbf{u'}}\right)\Delta \mathbf{u}\Delta \mathbf{u'},
\notag
\end{align}
where $b^\mathbf{j,k}$ is also an arbitrary function.
Therefore, the first point is automatically preserved if a finite-difference operator has a linearity.
In addition, the second point is naturally satisfied unless the Braams and Karney potential is employed.
The most important discussion in our article is the third point.
To satisfy $a^\mathbf{j,k}=0$ in the relativistic LFP equation,
the computational domain must be large enough to ensure that the distribution function
is negligible small at the boundary.
In other words, no boundary condition can enforce all of the conservation laws.

\subsection{\label{sec:3.1} MOMENTUM CONSERVATION}
The momentum of species ``$\mathrm{s}$'' is described as a first-order moment of Eq.~(\ref{eq:2.1}):

\begin{align}
&\frac{\partial}{\partial t}
\left(\sum_\mathbf{j}m_\mathrm{s}\mathbf{u^j}{f_\mathrm{s}}^\mathbf{j}\Delta \mathbf{u}\right)\notag\\
=&\frac{\Gamma}{2{m_\mathrm{s}}}\sum_\mathbf{j}\left[\mathbf{u^j}\frac{\delta}{\delta \mathbf{u}}\right.\cdot
\sum_\mathbf{k} \mathsf{U}(\mathbf{u^j},\mathbf{{u'}^k})\cdot\notag\\
&\qquad\qquad\left.\left({f_\mathrm{s'}}^\mathbf{k} \frac{\delta {f_\mathrm{s}}^\mathbf{j}}{\delta \mathbf{u}}
-\frac{m_\mathrm{s}}{m_\mathrm{s'}}{f_\mathrm{s}}^\mathbf{j} \frac{\delta {f_\mathrm{s'}}^\mathbf{k}}{\delta \mathbf{u'}}
\right)\Delta \mathbf{u'}\right]\Delta\mathbf{u}\notag\\
=&-\frac{\Gamma}{2}\sum_\mathbf{j,k}\left[\mathsf{U}(\mathbf{u^j},\mathbf{{u'}^k})\cdot
\left(\frac{{f_\mathrm{s'}}^\mathbf{k}}{m_\mathrm{s}} \frac{\delta {f_\mathrm{s}}^\mathbf{j}}{\delta \mathbf{u}}
-\frac{{f_\mathrm{s}}^\mathbf{j}}{m_\mathrm{s'}} \frac{\delta {f_\mathrm{s'}}^\mathbf{k}}{\delta \mathbf{u'}}
\right)\right]\Delta \mathbf{u}\Delta \mathbf{u'},\notag\\
&\left(\because \delta \mathbf{u^j}/\delta \mathbf{u}\mbox{ is an identity matrix.}\right)
\label{eq:3.1.1}
\end{align}
Likewise, the momentum of species ``$\mathrm{s'}$'' is obtained as follows:

\begin{align}
&\frac{\partial}{\partial t}
\left(\sum_\mathbf{k}m_\mathrm{s'}\mathbf{{u'}^k}{f_\mathrm{s'}}^\mathbf{k}\Delta \mathbf{u'}\right)\notag\\
=&-\frac{\Gamma}{2}\sum_\mathbf{j,k}\left[\mathsf{U}(\mathbf{{u'}^k},\mathbf{u^j})\cdot
\left(\frac{{f_\mathrm{s}}^\mathbf{j}}{m_\mathrm{s'}} \frac{\delta {f_\mathrm{s'}}^\mathbf{k}}{\delta \mathbf{u'}}
-\frac{{f_\mathrm{s'}}^\mathbf{k}}{m_\mathrm{s}} \frac{\delta {f_\mathrm{s}}^\mathbf{j}}{\delta \mathbf{u}}
\right)\right]\Delta \mathbf{u}\Delta \mathbf{u'}.
\label{eq:3.1.2}
\end{align}
The temporal development of total momentum is obtained as a sum of Eqs.~(\ref{eq:3.1.1}) and (\ref{eq:3.1.2}):

\begin{align}
&\frac{\partial}{\partial t}
\left(
\sum_\mathbf{j}m_\mathrm{s}\mathbf{{u}^j}{f_\mathrm{s}}^\mathbf{j}\Delta \mathbf{u}
+\sum_\mathbf{k}m_\mathrm{s'}\mathbf{{u'}^k}{f_\mathrm{s'}}^\mathbf{k}\Delta \mathbf{u'}
\right)\notag\\
=&-\frac{\Gamma}{2}\sum_\mathbf{j,k}\left[\left(\mathsf{U}(\mathbf{u^j},\mathbf{{u'}^k})
-\mathsf{U}(\mathbf{{u'}^k},\mathbf{{u}^j})\right)
\cdot\right.\notag\\
&\qquad\qquad\qquad\qquad
\left.\left(\frac{{f_\mathrm{s'}}^\mathbf{k}}{m_\mathrm{s}} \frac{\delta {f_\mathrm{s}}^\mathbf{j}}{\delta \mathbf{u}}
-\frac{{f_\mathrm{s}}^\mathbf{j}}{m_\mathrm{s'}} \frac{\delta {f_\mathrm{s'}}^\mathbf{k}}{\delta \mathbf{u'}}
\right)\right]\Delta \mathbf{u}\Delta \mathbf{u'}.\label{eq:3.1.3}
\end{align}
Therefore, the intuitive kernel of Eq.~(\ref{eq:2.2}) strictly maintains the momentum conservation owing to Eq.~(\ref{eq:2.3}).

\subsection{\label{sec:3.2} ENERGY CONSERVATION}
The energy of species ``$\mathrm{s}$'' is described as a second-order moment of Eq.~(\ref{eq:2.1}):

\begin{align}
&\frac{\partial}{\partial t}
\left(\sum_\mathbf{j}m_\mathrm{s}c^2\gamma^\mathbf{j}{f_\mathrm{s}}^\mathbf{j}\Delta \mathbf{u}\right)\notag\\
=&\frac{\Gamma}{2{m_\mathrm{s}}}\sum_\mathbf{j}\left[\gamma^\mathbf{j}c^2\frac{\delta}{\delta \mathbf{u}}\right.\cdot
\sum_\mathbf{k} \mathsf{U}(\mathbf{u^j},\mathbf{{u'}^k})\cdot\notag\\
&\qquad\qquad\left.\left({f_\mathrm{s'}}^\mathbf{k} \frac{\delta {f_\mathrm{s}}^\mathbf{j}}{\delta \mathbf{u}}
-\frac{m_\mathrm{s}}{m_\mathrm{s'}}{f_\mathrm{s}}^\mathbf{j} \frac{\delta {f_\mathrm{s'}}^\mathbf{k}}{\delta \mathbf{u'}}
\right)\Delta \mathbf{u'}\right]\Delta\mathbf{u}\notag\\
=&-\frac{\Gamma}{2}\sum_\mathbf{j,k}\left[\mathsf{U}(\mathbf{u^j},\mathbf{{u'}^k})\cdot
\frac{\delta (\gamma^\mathbf{j}c^2)}{\delta \mathbf{u}}\cdot\right. \notag\\
&\qquad\qquad\qquad
\left.\left(\frac{{f_\mathrm{s'}}^\mathbf{k}}{m_\mathrm{s}} \frac{\delta {f_\mathrm{s}}^\mathbf{j}}{\delta \mathbf{u}}
-\frac{{f_\mathrm{s}}^\mathbf{j}}{m_\mathrm{s'}} \frac{\delta {f_\mathrm{s'}}^\mathbf{k}}{\delta \mathbf{u'}}
\right)\right]\Delta \mathbf{u}\Delta \mathbf{u'}.\label{eq:3.2.1}
\end{align}
Likewise, the energy of species ``$\mathrm{s'}$'' is obtained as follows:

\begin{align}
&\frac{\partial}{\partial t}
\left(\sum_\mathbf{k}m_\mathrm{s'}c^2{\gamma'}^\mathbf{k}{f_\mathrm{s'}}^\mathbf{k}\Delta \mathbf{u'}\right)\notag\\
=&-\frac{\Gamma}{2}\sum_\mathbf{j,k}\left[\mathsf{U}(\mathbf{{u'}^k},\mathbf{{u}^j})\cdot
\frac{\delta ({\gamma'}^\mathbf{k}c^2)}{\delta \mathbf{u'}}\cdot\right. \notag\\
&\qquad\qquad\qquad
\left.\left(\frac{{f_\mathrm{s}}^\mathbf{j}}{m_\mathrm{s'}} \frac{\delta {f_\mathrm{s'}}^\mathbf{k}}{\delta \mathbf{u'}}
-\frac{{f_\mathrm{s'}}^\mathbf{k}}{m_\mathrm{s}} \frac{\delta {f_\mathrm{s}}^\mathbf{j}}{\delta \mathbf{u}}
\right)\right]\Delta \mathbf{u'}\Delta \mathbf{u}.\label{eq:3.2.2}
\end{align}

The temporal development of total energy is obtained as a sum of Eqs.~(\ref{eq:3.2.1}) and (\ref{eq:3.2.2}):

\begin{align}
&\frac{\partial}{\partial t}
\left(\sum_\mathbf{j}m_\mathrm{s}c^2\gamma^\mathbf{j}{f_\mathrm{s}}^\mathbf{j}\Delta \mathbf{u}
+\sum_\mathbf{k}m_\mathrm{s'}c^2{\gamma'}^\mathbf{k}{f_\mathrm{s'}}^\mathbf{k}\Delta \mathbf{u'}
\right)\notag\\
=&-\frac{\Gamma}{2}\sum_\mathbf{j,k}\left[\left(\mathsf{U}(\mathbf{u^j},\mathbf{{u'}^k})\cdot\mathbf{\bar{v}}
-\mathsf{U}(\mathbf{{u'}^k},\mathbf{{u}^j})\cdot\mathbf{\bar{v}'}\right)
\cdot\right.\notag\\
&\qquad\qquad\qquad\qquad
\left.\left(\frac{{f_\mathrm{s'}}^\mathbf{k}}{m_\mathrm{s}} \frac{\delta {f_\mathrm{s}}^\mathbf{j}}{\delta \mathbf{u}}
-\frac{{f_\mathrm{s}}^\mathbf{j}}{m_\mathrm{s'}} \frac{\delta {f_\mathrm{s'}}^\mathbf{k}}{\delta \mathbf{u'}}
\right)\right]\Delta \mathbf{u}\Delta \mathbf{u'},\label{eq:3.2.3}
\end{align}
where the velocities with overlines are defined as follows:

\begin{align}
\mathbf{\bar{v}}\equiv \frac{\delta (\gamma^\mathbf{j}c^2)}{\delta \mathbf{u}},\quad
\mathbf{\bar{v}}'\equiv \frac{\delta ({\gamma'}^\mathbf{k}c^2)}{\delta \mathbf{u'}}.\label{eq:3.2.4}
\end{align}
Therefore,
$\mathsf{U}(\mathbf{u^j},\mathbf{{u'}^k})\cdot\mathbf{\bar{v}}=\mathsf{U}(\mathbf{{u'}^k},\mathbf{u^j})\cdot\mathbf{\bar{v}'}$
is required for the energy conservation in discrete form.
The intuitive kernel Eq.~(\ref{eq:2.2}) satisfies Eq.~(\ref{eq:2.4}),
so the energy conservation would be preserved even in discrete form if
$\mathbf{\bar{v}}=\mathbf{v^j}$ and $\mathbf{\bar{v}'}=\mathbf{{v'}^k}$ were true.
Generally speaking, the proposition is false resulting in a violation of the energy conservation.
For the second-order central difference;

\begin{align}
(\gamma^{j_1+1,j_2,j_3})^2-(\gamma^{j_1-1,j_2,j_3})^2=({u_1}^{j_1+1}/c)^2-({u_1}^{j_1-1}/c)^2,\notag\\
\therefore \bar{v}_1=\frac{\delta (\gamma^\mathbf{j}c^2)}{\delta u_1}
=\frac{2{u_1}^{j_1}}{\gamma^{j_1+1,j_2,j_3}+\gamma^{j_1-1,j_2,j_3}}
\neq\frac{{u_1}^{j_1}}{\gamma^{j_1,j_2,j_3}}.\notag
\end{align}
The only exception is the non-relativistic limit when the Lorentz factor is always unity.
Hirvijoki and Adams reported that their scheme cannot maintain the exact energy conservation in the relativistic regime
\cite{Hirvijoki2017}, and our discussion should be connected on the fundamental level with this issue
although their discussion was based on the finite-element method.

\section{\label{sec:4}STRUCTURE-PRESERVING STRATEGY}

Here we propose a structure-preserving scheme for the relativistic LFP equation
which resolves the energy-conservation problem of the intuitive scheme.
According to Eqs.~(\ref{eq:3.1.3}) and (\ref{eq:3.2.3}), the following discrete requirements must be preserved
to maintain the mass-momentum-energy conservation:

\begin{enumerate}
   \item Summation-by-parts
   \item $\mathsf{U}(\mathbf{u^j},\mathbf{{u'}^k})=\mathsf{U}(\mathbf{{u'}^k},\mathbf{u^j})$
   \item $\mathsf{U}(\mathbf{u^j},\mathbf{{u'}^k})\cdot\mathbf{\bar{v}}=\mathsf{U}(\mathbf{{u'}^k},\mathbf{u^j})\cdot\mathbf{\bar{v}'}$
\end{enumerate}
The following is the only Beliaev--Budker kernel that preserves the above requirements:

\begin{align}
\mathsf{U}(\mathbf{u^j},\mathbf{u'}^\mathbf{k})\equiv\frac{\bar{r}^2}{\bar{\gamma}\bar{\gamma}' \bar{w}^3}
\left[\bar{w}^2\mathsf{I}-\bar{\gamma}^2\mathbf{\bar{v}\bar{v}}-\bar{\gamma}'^2\mathbf{\bar{v}'\bar{v}'}\right.\notag\\
\left.+\bar{r}\bar{\gamma}\bar{\gamma}'(\mathbf{\bar{v}\bar{v}'+\bar{v}'\bar{v}})\right],
\label{eq:4.1}
\end{align}
where the variables with overlines are

\begin{align}
\bar{\gamma}\equiv 1/\sqrt{1-|\bar{\mathbf{v}}/c|^2}\neq\gamma^\mathbf{j},\quad
\bar{\gamma}'\equiv 1/\sqrt{1-|\bar{\mathbf{v}}'/c|^2}\neq{\gamma'}^\mathbf{k},\notag\\
\bar{r}\equiv \bar{\gamma}\bar{\gamma}'(1-\bar{\mathbf{v}}\cdot\bar{\mathbf{v}}')\neq r^\mathbf{j,k},\quad
\bar{w}\equiv c\sqrt{\bar{r}^2-1}\neq w^\mathbf{j,k}.\notag
\end{align}
A combination of Eqs.~(\ref{eq:2.1}) and (\ref{eq:4.1}) naturally preserves the law of energy conservation naturally.
However, the positivity of distribution function and H-theorem are not guaranteed unconditionally by this formulation.
We do not give a discussion of the temporal discretization in this article,
which is done in the papers about entropic schemes.

\section{\label{sec:5}DEMONSTRATION}

\begin{figure}
\includegraphics[width=0.5\textwidth]{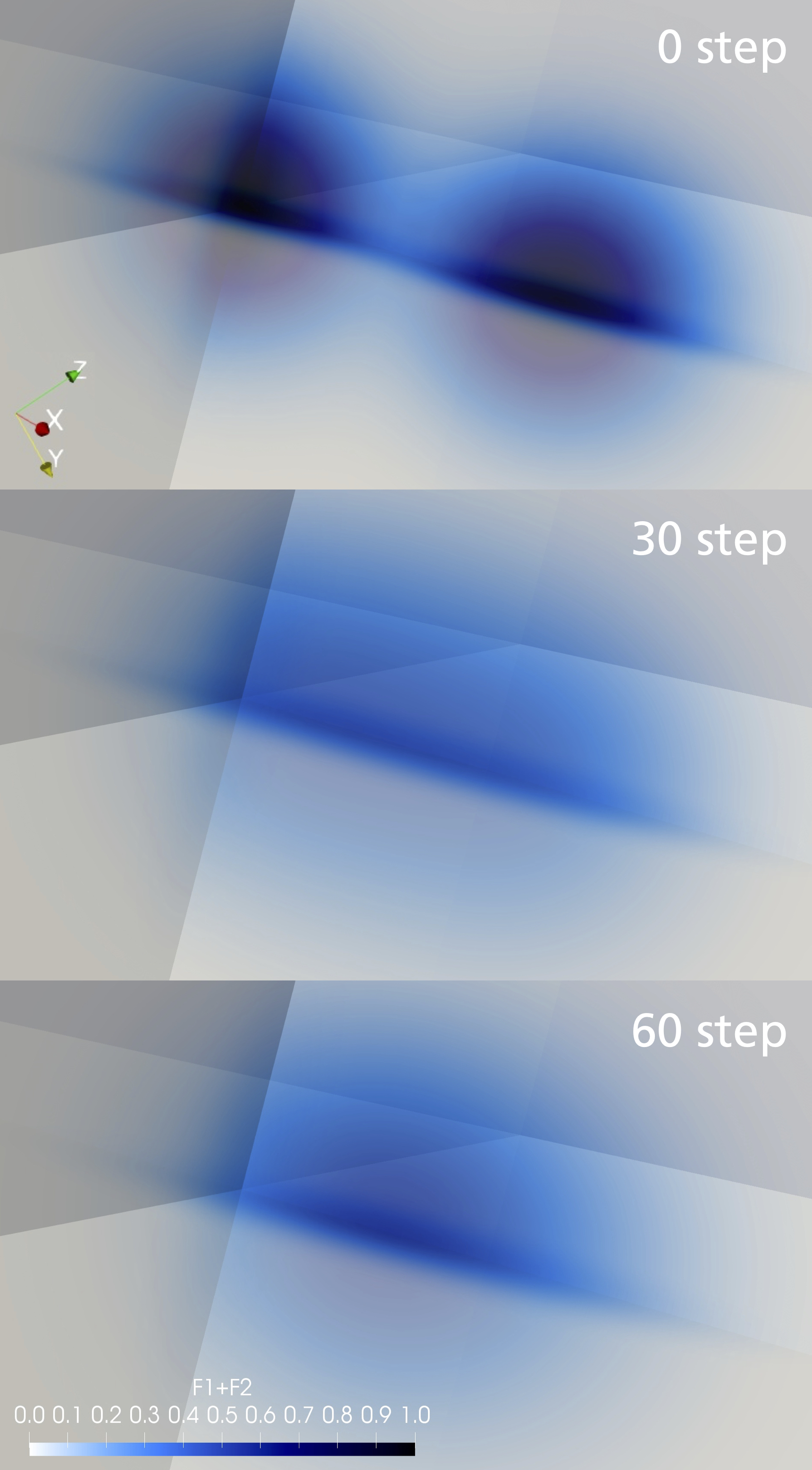}
\caption{\label{fig:1} (Color online) Time development of distribution function
initialized by double Maxwell--J\"{u}ttner distribution.
}
\end{figure}

\begin{figure}
\subfigure[Structure-preserving discretization (proposed).]
{\includegraphics[width=0.5\textwidth]{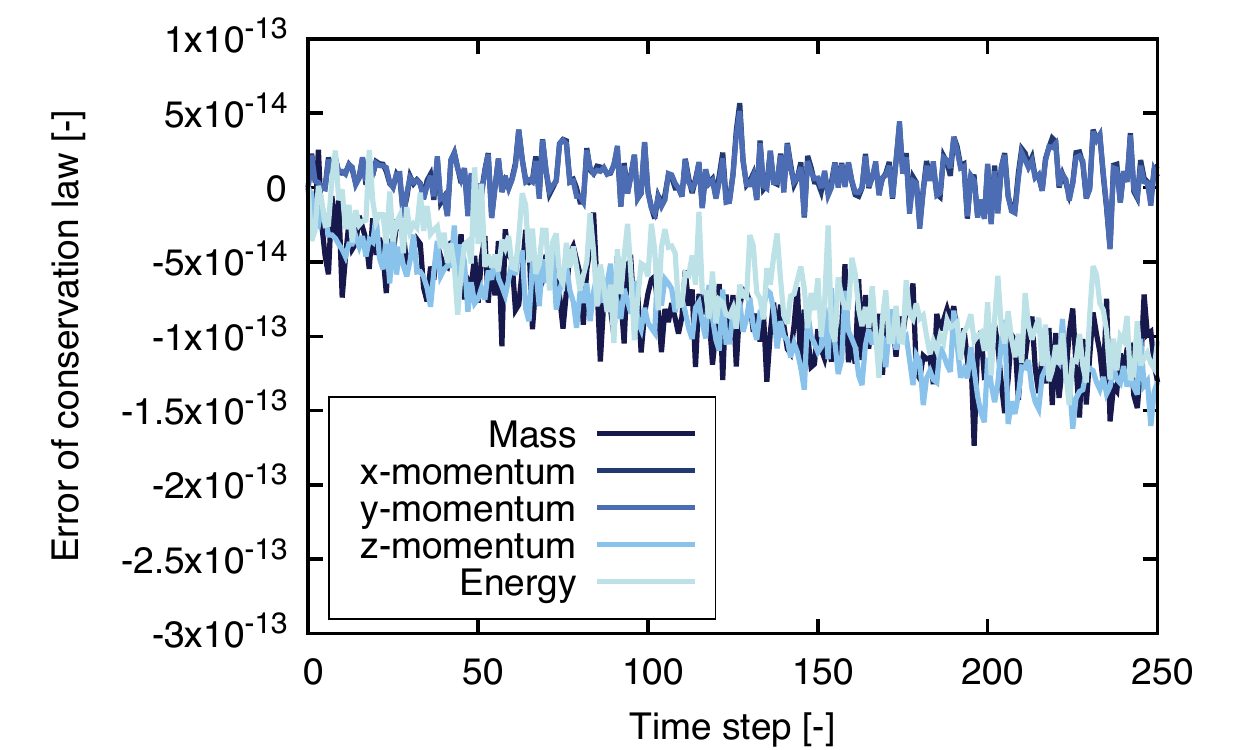}}
\subfigure[Intuitive discretization (conventional).]
{\includegraphics[width=0.5\textwidth]{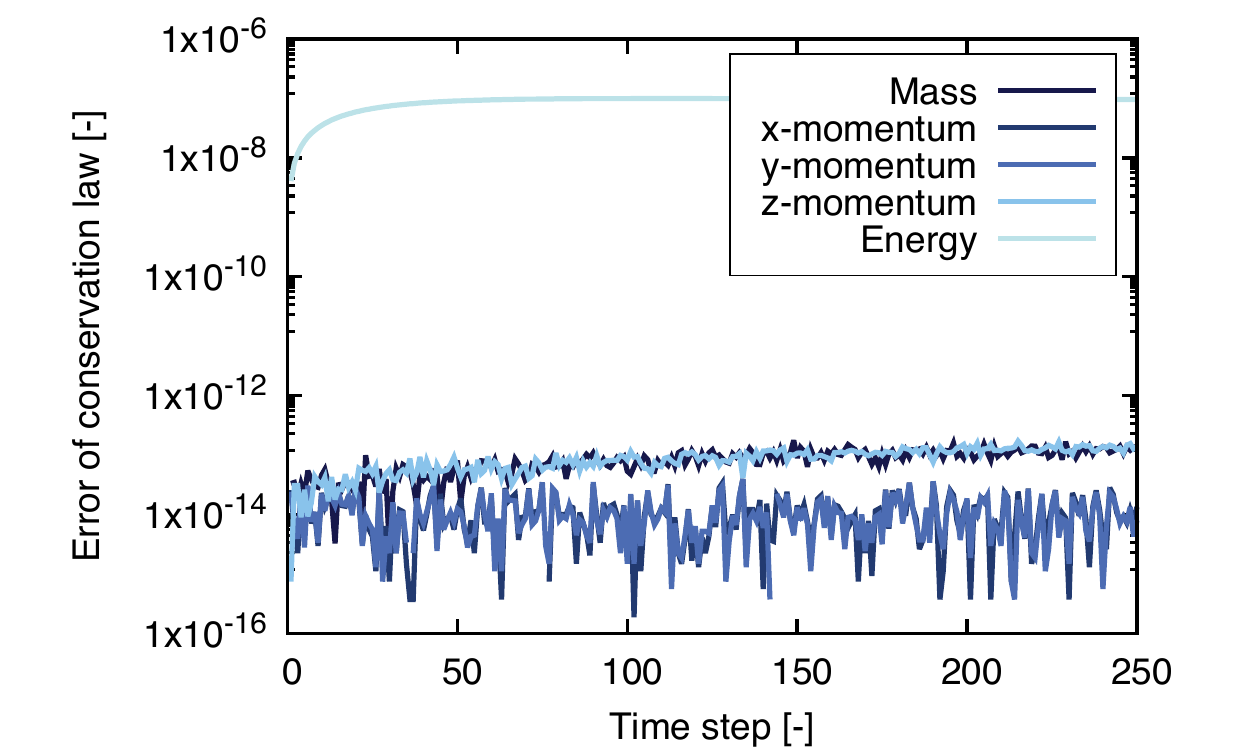}}
\caption{\label{fig:2} Conservation errors for mass, momentum and energy in a thermal-equilibration verification.}
\end{figure}

\begin{figure}
\includegraphics[width=0.5\textwidth]{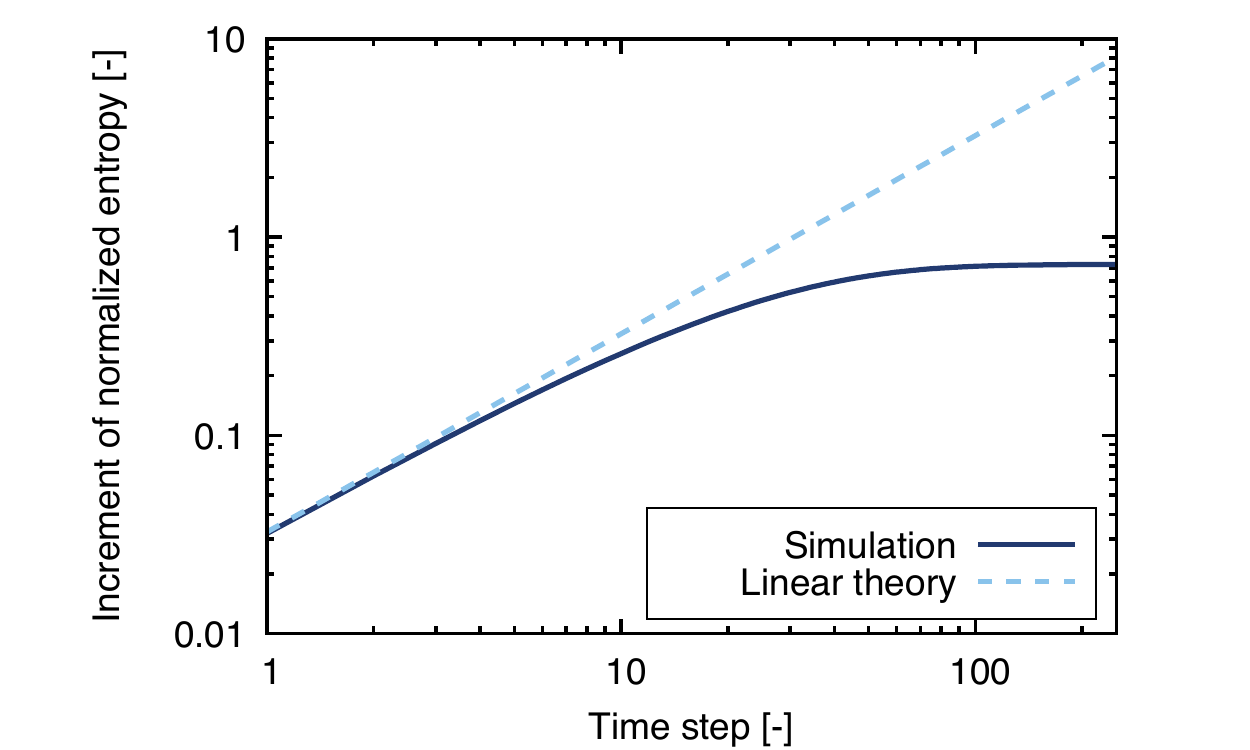}
\caption{\label{fig:4} (Color online) Verification of H-theorem.
Caution: the proposed scheme does not ensure the H-theorem mathematically.
}
\end{figure}

To verify the proposed scheme, a collisional relaxation of a particle--antiparticle plasma
is calculated for simplicity ($m_\mathrm{s}=m_\mathrm{s'}\equiv m, q_\mathrm{s}=-q_\mathrm{s'}\equiv q$).
The particles and antiparticles are initialized by the following shifted Maxwell--J\"{u}ttner distrubtions \cite{Zenitani2015}.

\begin{align}
f_\mathrm{s}(\mathbf{u})=\exp\left(-\frac{\gamma-1}{\theta}\right),\notag\\
f_\mathrm{s'}(\mathbf{u'})=\frac{1}{\gamma_0}\exp\left(-\frac{\gamma'\gamma_0-\gamma_0\mathbf{v}_0\cdot\mathbf{u'}/c^2-1}{\theta}\right),\notag
\end{align}
where $\theta= 0.01$ is the temperature normalized by the rest mass energy,
$\mathbf{v}_0/c=^\mathrm{T}[0.2,0.2,0.2]$, and $\gamma_0=1/\sqrt{1-\mathbf{v}_0\cdot \mathbf{v}_0/c^2}$.
The computational domain is set to be $\{(u_1,u_2,u_3)|-1.2c\le u_1,u_2,u_3\le 1.8c\}$,
and the number of computational cells is $128\times128\times128$.
In this verification, a first-order Euler explicit method is used as a time integration.
The temporal interval is given as $\Gamma m^{-2}\Delta t=1/80$,
so that the temporal resolution is fine enough to see the relaxation.
In this test, only unlike-particle collisions are considered.

Figure~\ref{fig:1} shows a time development of the double Maxwell--J\"{u}ttner distribution through the collisional relaxation.
The double-peaked distribution function approximates to the equilibrium state whose velocity shift and temperature are
$\mathbf{v}_0/2$ and $2\theta$, respectively.
Figures~\ref{fig:2}(a) and (b) indicate the errors of conservation laws for the structure-preserving and
intuitive discretizations, respectively.
The structure-preserving scheme with the collision kernel of Eq.~(\ref{eq:4.1}) strictly preserves
the conservation laws of mass, momentum and energy only with round-off errors.
It seems that the mass, momentum of $z$-direction and energy slightly accumulate the round-off errors.
We reported that Pad\'{e}-type (or implicit) filters used in computational fluid dynamics can
accumulate round-off errors, but it can be suppressed by changing the order of arithmetic operations \cite{Shiroto2017}.
Therefore, the accumulation of round-off errors can be resolved by an optimization of the code
although it is out of scope of this article.
From the other perspective, the conservation property of the proposed scheme is quite fine
as well as it is affected by the order of arithmetic operations.
In contrast, the intuitive discretization with the collision kernel of Eq.~(\ref{eq:2.2}) clearly violates the energy conservation.
Note that the mass and momentum are strictly conserved with the intuitive scheme.
This is because the relativistic LFP equation is discretized in the divergence-form,
and the intuitive collision kernel also satisfies
the symmetry of $\mathsf{U}(\mathbf{u},\mathbf{u'})=\mathsf{U}(\mathbf{u'},\mathbf{u})$.

Finally, we address some remaining issues about the positivity and H-theorem.
Figure~\ref{fig:4} shows the H-theorem is strictly preserved and
the linear relaxation rate is initially reproduced well in the current situation.
Here we introduced the entropy as follows to avoid negative values due to round-off errors:

\begin{align}
S\equiv -\sum_\mathbf{j} |{f_\mathrm{s}}^\mathbf{j}|\log|{f_\mathrm{s}}^\mathbf{j}|\Delta \mathbf{u}
-\sum_\mathbf{k} |{f_\mathrm{s'}}^\mathbf{k}|\log|{f_\mathrm{s'}}^\mathbf{k}|\Delta \mathbf{u'}.\notag
\end{align}
The linear relaxation rate is obtained as follows by assuming the initial Maxwellian:

\begin{align}
\left(\frac{\Gamma}{2m^2}\right)^{-1}\frac{1}{S(0)}\frac{\mathrm{d}S}{\mathrm{d}t}\simeq 5.2.\notag
\end{align}
Reference~\cite{Berezin1987}, for example, suggests that there is an upper-bound of $\Delta t$
which ensures the positivity and H-theorem.
In this experiment, the collision time is resolved quite well so that the requirement is satisfied.
However, such a small temporal interval is not suit for practical simulations.
We did not give a discussion about the temporal discretization in this article
since the conservation laws do not depend on the temporal structure of the relativistic LFP equation.
A development of conservative and entropic scheme for the relativistic LFP equation will be performed
in the separate paper.

\section{\label{sec:6}CONCLUSIONS}
A feasibility of conservative scheme for the relativistic Landau--Fokker--Planck equation
has been demonstrated.
The proposed scheme has a unique way of calculating the collision kernel specialized for linear finite-difference operators.
The verification via thermal-equilibration problem manifests the conservation of mass, momentum and energy only with round-off errors.
Although there are still some problems of computational cost, positivity, H-theorem and boundary conditions,
our strategy gives a piece of puzzle for practical simulation of the collisional relaxation in the relativistic regime.

\begin{acknowledgments}
T.S. would like to appreciate valuable information from Dr. William T. Taitano and Dr. Luis Chac\'{o}n
(Los Alamos National Laboratory). This work was supported by KAKENHI Grant Numbers JP15K21767 and JP18H05851.
Numerical experiments were carried out on NEC SX-ACE, Cybermedia Center, Osaka University.
\end{acknowledgments}

%
%

\nocite{*}
\bibliography{aipsamp}

\end{document}